# A Physics-informed machine learning model for time-dependent wave runup prediction

Saeed Saviz Naeini[a], Reda Snaiki[a,*]

[a] Department of Construction Engineering, École de Technologie Supérieure, Université du Québec, Montréal, QC, H3C 1K3, Canada

[*]*Corresponding author. Email:* reda.snaiki@etsmtl.ca

**Abstract**: Wave runup is a critical factor affecting coastal flooding, shoreline changes, and damage to coastal structures. Climate change is also expected to amplify wave runup's impact on coastal areas. Therefore, fast and accurate wave runup estimation is essential for effective coastal engineering design and management. However, predicting the time-dependent wave runup is challenging due to the intrinsic nonlinearities and non-stationarity of the process, even with the use of the most advanced machine learning techniques. In this study, a physics-informed machine learning-based approach is proposed to efficiently and accurately simulate time-series wave runup. The methodology combines the computational efficiency of the Surfbeat (XBSB) mode with the accuracy of the nonhydrostatic (XBNH) mode of the XBeach model. Specifically, a conditional generative adversarial network (cGAN) is used to map the image representation of wave runup from XBSB to the corresponding image from XBNH. These images are generated by first converting wave runup signals into time-frequency scalograms and then transforming them into image representations. The cGAN model achieves improved performance in image-to-image mapping tasks by incorporating physics-based knowledge from XBSB. After training the model, the high-fidelity XBNH-based scalograms can be predicted, which are then employed to reconstruct the time-series wave runup using the inverse wavelet transform. The simulation results underscore the efficiency and robustness of the proposed model in predicting wave runup, suggesting its potential value for applications in risk assessment and management.

## 1. Introduction

The phenomenon of storm-induced wave runup poses significant risks of erosion, flooding, and damage to coastal infrastructures. Several factors affect the wave runup magnitude including the wave characteristics, coastal topography, and the presence of engineered structures such as breakwaters (Muttray et al., 2007; Roelvink et al., 2009). As sea levels continue to rise and storm intensities increase as a result of climate change, wave runup is becoming an increasingly important issue since it is expected to exacerbate the effects of flooding events along coastal areas (Wolf, 2009; Didier et al., 2015; Snaiki et al., 2020; Safari Ghaleh et al., 2021). Consequently, accurate modeling of wave runup is crucial to support coastal engineers in designing and managing coastal infrastructure effectively to mitigate the adverse effects of extreme storm events (Ruggiero et al., 2001; Casella et al., 2014; Kijewski-Correa et al., 2020; Hermawan et al., 2023). Additionally, by understanding and simulating the fundamental mechanisms of wave runup, stakeholders and planners can make judicious decisions to enhance coastal resilience, protect vulnerable areas, and minimize the potential risks posed by storm-induced flooding.

Several approaches have been developed to simulate the wave runup mechanism, including the numerical models. The latter can be generally classified into phase-averaged and phase-resolving numerical approaches (Fiedler et al., 2018). While phase-resolving models effectively capture individual wave phases, leading to a higher level of precision in simulating the wave characteristics, they come with a higher computational cost. On the other hand, phase-averaged models are faster and less accurate, as they provide an average representation of wave behavior over time intervals (Buckley et al., 2014; Lashley et al., 2018). For instance, the Boussinesq equations, which are derived from the Navier-Stokes equations, can resolve the wave properties and capture both the long-term effects of wave setup and the short-term effects of wave breaking (Chen et al., 2000; Kennedy et al., 2000). Several numerical tools and software applications have been developed based on the Boussinesq equations to simulate the wave runup, such as BOSZ, FUNWAVE, BOUSS-2D, BOUSS-1D and XBeach (Kennedy et al., 2012; Stockdon et al., 2014; Quataert et al., 2015; Pearson et al., 2017; Lashley et al., 2018; Roelvink et al., 2018; Yao et al., 2018; Ning et al., 2019a; Ning et al., 2019b; Pinault et al., 2020; Rutten et al., 2021; Li et al., 2021; Amini and Marsooli, 2023). For example, Hatzikyriakou and Lin (2018) assessed the vulnerability of structures and residential communities to coastal flooding from Hurricane Sandy along the coastlines of New York and New Jersey. The initial step involved simulating the storm surge and waves over the Atlantic Ocean, spanning six days prior to and one day after Sandy's landfall using the coupled Advanced CIRCulation (ADCIRC) and Simulating WAves Nearshore (SWAN) numerical models. The outputs from the ADCIRC and SWAN models were subsequently utilized to simulate the wave runup, wave overtopping and inland flooding. This simulation employed the one-dimensional phase-resolving Boussinesq surf zone model (BOUSS1D), which was implemented using a sequence of 1D cross-shore transects with a 2 m grid spacing. On the other hand, De Beer et al. (2021) applied the short-wave resolving, Nonhydrostatic (XBNH) and short-wave averaged, Surfbeat (XBSB) modes of the XBeach numerical model to simulate wave runup and swash on an intermediate-reflective sandy beach in Duck, North Carolina. It was concluded that XBNH effectively models the incident-band swash, infragravity-band swash, and runup in comparison to the XBSB which underestimates the incident-band swash and, to lesser extent, the infragravity-band swash.

In scenarios involving probabilistic analysis and risk assessment (Snaiki and Parida, 2023a, 2023b), the utilization of high-fidelity numerical models may not prove efficient. Hence, alternative models, including empirical and machine learning techniques, have been devised to rapidly estimate wave runup in terms of important factors such as the wave height, wave period, beach slope and other beach characteristics (Didier et al., 2016; Vinodh and Tanaka, 2020). For example, the empirical models are usually based on simplified mathematical formulations to estimate some statistical parameters, including the $R_{max}$ and $R_{2\%}$ which represent the highest runup achieved at any given time and the 2% wave runup level (i.e., the elevation exceeded by only 2% of the runup events), respectively (e.g., Hunt, 1959; Holman, 1986; Mase, 1989; Van der Meer and Stam, 1992; Hedges and Mase, 2004; Ruggiero et al., 2004; Stockdon et al., 2006; Vousdoukas et al., 2012; Park and Cox, 2016). While several parameterizations have been developed for the empirical models, various formulations expressed the runup level in terms of the nondimensional Iribarren number, also known as the surf similarity parameter, which accounts simultaneously for beach and wave characteristics (Senechal et al., 2011; Blenkinsopp et al., 2016). The empirical models are usually developed based on numerical or observed data from field measurements or laboratory experiments (Atkinson et al., 2017). Despite their simplicity and practicality, these models are constrained by various limitations, as the simplified formulations might fail to capture the inherent nonlinearities within the model. Furthermore, the effectiveness of these models can vary depending on the particular coastal conditions and wave characteristics being considered. Moreover, they might have limitations in accurately predicting wave runup during extreme conditions (Guimaraes et al., 2015; Cohn and Ruggiero, 2016; Di Luccio et al., 2018). Therefore, alternative approaches based on advanced data driven techniques (e.g., machine learning) have been recently developed to overcome these limitations (Wu and Snaiki, 2022). For example, Power et al. (2019) compared the predictive ability of seven empirical models with the Gene-Expression Programming (GEP) model for the estimation of $R_{2\%}$. The training/testing data were collected from both field and laboratory measurements corresponding to various beach configurations involving different sediment sizes and bed roughness. The obtained results revealed that the GEP model successfully captured the nonlinear effects of the wave runup mechanism and hence outperformed all other empirical models. Rehman et al. (2022) predicted $R_{max}$ over different arrays of rubble mound and caisson-type breakwaters using two techniques, namely a feed-forward artificial neural network (ANN) and a response surface methodology (RSM). The training dataset was retrieved from numerous experimental tests conducted within a laboratory flume. Although both ANN and RSM models effectively simulated the maximum wave runup, the statistical performance of the former exhibited a slight superiority. Tarwidi et al. (2023) trained an extreme gradient boosting (XGBoost) model based on experimental datasets to estimate the relative wave runup on a sloping beach. Furthermore, the XGBoost model was compared against three machine learning models, namely the multiple linear regression, support vector regression, and random forest. The XGBoost model demonstrated superior performance compared to all other machine learning methods in predicting wave runup. While various alternative machine learning approaches have been employed to model the wave runup (e.g., Bonakdar and Etemad-Shahidi, 2011; Beuzen et al., 2019; Yao et al., 2021; Mahdavi-Meymand et al., 2022), most of these applications focus on predicting a limited set of statistical measures such as $R_{max}$ and $R_{2\%}$.

Given the intrinsic nonlinearities of the wave runup process, coupled with its transient and nonstationary characteristics, predicting the temporal evolution of wave elevation remains exceptionally intricate, even with the application of cutting-edge machine learning techniques. As a result, innovative strategies are needed to assist machine learning methods in tackling these challenges. This study introduces a novel physics-informed machine learning-based approach for simulating time-series wave runup which combines the efficiency of XBSB with the accuracy of XBNH mode. The training datasets are generated from both XBSB and XBNH wave runup simulations, corresponding to several storm scenarios. These scenarios are implemented as boundary conditions within the selected basin using a simplified representation based on the JONSWAP spectrum. The proposed model takes the wave runup values generated through the XBSB mode as input and predicts the corresponding wave runup values using the XBNH mode. Specifically, the signals generated from both the XBSB and XBNH models are initially transformed into the time-frequency domain using the Morlet wavelet transform technique, resulting in scalograms. The obtained scalograms are subsequently converted into images and used by a Conditional Generative Adversarial Network (cGAN) which is employed to establish a mapping between the image representations of the scalograms derived from the XBSB and XBNH modes. Once the model is trained, the high-fidelity XBNH-based scalograms can be predicted which are subsequently used to reconstruct the time-series wave runup results through the inverse wavelet transform. The proposed methodology is demonstrated using a simplified case study involving a 1D basin profile with varying depths. In addition, the performance of the proposed model in predicting wave runup is assessed across multiple case scenarios.

## 2. Numerical Prediction of Wave Runup

### 2.1. Wave runup simulation overview

Wave runup refers to the maximum onshore elevation reached by waves on a beach or shoreline. It is an important component which affects the coastal inundation and coastal erosion (Stockdon et al., 2006; Roelvink et al., 2009). During extreme conditions (e.g., hurricanes) (Snaiki and Wu, 2020a), the combination of high waves, storm surge, high tides and currents has a significant impact on wave runup which will ultimately determine the extent of the induced flooding (Kennedy et al., 2012; Marcos et al., 2019). The main factors affecting wave runup include the wave height, wave period, beach slope, bed roughness, tidal level, and wind set-up. The most common methods to simulate wave runup include the empirical models, the numerical models and the physical (experimental) models (Li and Raichlen, 2003; Liang et al., 2013; Guimaraes et al., 2015). Although, physical models are useful for understanding the physical processes involved in wave runup, they are usually expensive and time-consuming to set up. The empirical models can only be used for simple applications since they cannot capture the full complexity of wave runup. Numerical models, on the other hand, can solve complex fluid dynamics problems with high accuracy. To simulate wave runup on coastal areas, several numerical models have been developed. For example, ADCIRC is usually coupled with the SWAN model to simulate the storm surge and waves for shallow waters (Luettich et al., 1992; Booij et al., 1999). However, the high computational requirements of these models limit their ability to resolve small-scale coastal features (Saviz Naeini and Snaiki, 2023). As a result, these models may be accurate for simulating large-scale processes, but may not be able to capture the intricate details of near-shore dynamics (Hatzikyriakou and Lin, 2018). To overcome these limitations, Boussinesq equations have been proposed to simulate the propagation and transformation of waves in shallow water, as well as to estimate the runup height and velocity of waves as they approach the coastline. These equations are derived from the Navier-Stokes equations and make simplifying assumptions, such as neglecting the vertical acceleration of the water and averaging the velocity field over a vertical layer. This allows them to resolve small-scale coastal characteristics that would be difficult to capture with other models (Yao et al., 2012; Su et al., 2015). Several numerical models and software have been developed based on the Boussinesq equations, including FUNWAVE, BOSZ, BOUSS-2D, BOUSS-1D and XBeach. The latter is one of the most widely used and recognized open-source models for simulating wave runup and other coastal processes. It has been rigorously applied and validated to simulate a wide range of wave phenomena, including wave propagation, wave breaking, and wave runup (Roelvink et al., 2009, 2018). Due to its robustness, XBeach model will be used in this study for the simulation of wave runup. Further discussion about the XBeach model is provided in the next section.

## 2.2. XBeach model overview

XBeach is a numerical model that simulates hydrodynamic and morphodynamic processes on coastlines. It offers three distinct modes, each with different levels of computational efficiency and complexity, namely the Stationary mode, the Surf-Beat (XBSB) mode and the Nonhydrostatic (XBNH) mode (Bart, 2017; Ruffini et al., 2020). The Stationary mode is the simplest model in XBeach since it solves the wave-averaged equations, focusing primarily on short waves and neglecting infragravity waves. This mode is computationally efficient and suitable for scenarios where the long-wave effects can be neglected. The XBSB mode, also denoted as the instationary mode, is more accurate than the Stationary mode because it averages the short waves (short wave envelope) and resolves the long waves while accounting for the interaction between them. The XBNH mode is the most accurate mode, as it resolves both short and long waves by solving a combination of non-linear shallow water equations with a pressure correction term. This mode provides a comprehensive representation of wave dynamics but is computationally expensive, making it unsuitable for real-time risk evaluation (De Alegria-Arzaburu et al., 2011; Harris et al., 2018; Roelvink and Costas, 2019). Since both the XBSB and XBNH modes will be used in this study, a brief description of their mathematical formulation is presented below.

### 2.2.1 XBSB mode

The XBSB mode solves the short-wave averaged wave action balance equations with time-dependent forcing. This mode also employs a roller model to represent the momentum stored at the water surface after wave breaking (Svendsen, 1984; Nairn et al., 1991). The wave action balance equation used in the XBSB mode is given by:

$$\frac{\partial A}{\partial t} + \frac{\partial c_x A}{\partial x} + \frac{\partial c_y A}{\partial y} + \frac{\partial c_\theta A}{\partial \theta} = -\frac{D_w}{\sigma} \tag{1}$$

with the wave action:

$$A(x, y, t, \theta) = \frac{S_w(x,y,t,\theta)}{\sigma(x,y,t)} \tag{2}$$

where $S_w$ = wave energy density; $\sigma$ = intrinsic wave frequency; $h$ = local water depth; $k$ = wave number; $c_x, c_y$ = wave-action propagation speed in the $x$ and $y$ directions, respectively; $c_\theta$ = wave-action propagation speed in the directional $\theta$ space; and $D_w$ = dispersion due to wave breaking. The dispersion due to wave breaking can be modeled using several variations of the formulation proposed by (Roelvink, 1993). For example, it can be expressed as:

$$D_w(x, y, t, \theta) = \frac{S_w(x,y,t,\theta)}{E_w(x,y,t)} \left(2\frac{\alpha}{T_{rep}} Q_b E_w\right) \tag{3}$$

$$Q_b = 1 - \exp\left(-\left(\frac{H_{rms}}{H_{max}}\right)^n\right), \; H_{rms} = \sqrt{\frac{8E_w}{\rho g}}, \; H_{max} = \gamma \cdot (h + \delta H_{rms}) \tag{4}$$

where $\alpha$ = calibration coefficient for dissipation; $T_{rep}$ = representative wave period; $E_w$ = total wave energy; $H_{rms}$ = root-mean-square wave height; $\rho$ = water density; $g$ = gravitational acceleration; $H_{max}$ = maximum wave height; $n$ = empirical coefficient; $\gamma$ = breaker parameter; and $\delta H_{rms}$ = fraction of the wave height. The radiation stresses can be calculated using the linear wave theory, given the distribution of wave action and wave energy in space:

$$S_{xx,w}(x, y, t) = \int \left(\frac{c_g}{c}(1 + \cos^2 \theta) - \frac{1}{2}\right) S_w d\theta \tag{5}$$

$$S_{xy,w}(x, y, t) = S_{yx,w} = \int \sin \theta \cos \theta \left(\frac{c_g}{c} S_w\right) d\theta \tag{6}$$

$$S_{yy,w}(x, y, t) = \int \left(\frac{c_g}{c}(1 + \sin^2 \theta) - \frac{1}{2}\right) S_w d\theta \tag{7}$$

where $S_{xx,w}, S_{yy,w}, S_{xy,w}, S_{yx,w}$ = shear components of the radiation stress; $c_g$ = group velocity obtained from linear theory; $c$ = phase velocity; and $\theta$ = angle of incidence with respect to the $x$-axis. The generated surface wave stresses are subsequently transmitted to the long-wave model, which is based on nonlinear shallow water equations. This model is utilized to simulate the currents and long waves, such as surges. It is important to note that in the XBSB mode, the nonlinear shallow water equations solely consider the hydrostatic pressure.

### 2.2.2 XBNH mode

The XBNH mode is a phase-resolving model which resolves both the incident and infragravity wave motion. The depth-averaged flow is computed using the non-hydrostatic nonlinear shallow water equations which can be expressed for the two-dimensional case as:

$$\frac{\partial \eta}{\partial t} + \frac{\partial uh}{\partial x} + \frac{\partial vh}{\partial y} = 0 \tag{8}$$

$$\frac{\partial u}{\partial t} + u\frac{\partial u}{\partial x} + v\frac{\partial u}{\partial y} - v_h\left(\frac{\partial^2 u}{\partial x^2} + \frac{\partial^2 u}{\partial y^2}\right) = -\frac{\tau_{bx}}{\rho h} - \frac{1}{\rho}\frac{\partial(\bar{q}+\rho g \eta)}{\partial x} \tag{9}$$

$$\frac{\partial v}{\partial t} + u\frac{\partial v}{\partial x} + v\frac{\partial v}{\partial y} - v_h\left(\frac{\partial^2 v}{\partial x^2} + \frac{\partial^2 v}{\partial y^2}\right) = -\frac{\tau_{by}}{\rho h} - \frac{1}{\rho}\frac{\partial(\bar{q}+\rho g \eta)}{\partial y} \tag{10}$$

where $t$ = temporal coordinate; $u$ = depth-averaged velocity in $x$ (cross-shore) direction; $v$ = depth-averaged velocity in $y$ (alongshore) direction; $v_h$ = horizontal viscosity; $\rho$ = water density; $\bar{q}$ = depth-averaged dynamic nonhydrostatic pressure; $h$ = local water depth; and $\tau_{bx}, \tau_{by}$ = bed shear stresses in $x$ and $y$ direction. The XBNH mode is significantly more computationally demanding than the XBSB mode because it resolves both the long and short waves. This requires high spatial resolution and smaller time steps, which increases the computational cost.

### 3. Proposed Wave Runup Model

### 3.1 Proposed model overview

As mentioned earlier, both XBSB and XBNH have their own strengths and limitations. XBSB stands out for its relatively low computational requirements, making it suitable for various applications like probabilistic and risk assessment. However, its simplified assumptions, such as wave averaging, prevents it from achieving the same level of accuracy as XBNH. On the other hand, XBNH provides more precise simulations but entails high computational costs due to the involvement of multiple nonlinear equations. Thus, the objective of the proposed study is to develop a new model that leverages the efficiency of XBSB along with the accuracy of XBNH mode. Specifically, a machine learning model will be developed to predict the time-series wave runup. This model will take as input the wave runup simulations generated based on the XBSB mode and will predict the corresponding wave runup simulations from the XBNH mode. The machine learning model will be trained on a dataset generated from XBSB and XBNH wave runup simulations corresponding to several storm scenarios. Once the model is trained, it can be used to predict the wave runup for any given set of input conditions.

      The storm scenarios are first implemented as the boundary conditions of the selected basin using a simplified representation based on the JONSWAP spectrum. Subsequently, both the XBSB and XBNH models are run using these scenarios to generate the required database for model training and testing. The simulation results from the XBSB model are utilized as inputs for the machine learning model, while the simulation results from the XBNH model serve as the model's outputs. Predicting the time series wave runup, with its high fluctuations, is a highly intricate task.

Even with advanced algorithms like the long short-term memory networks (LSTM), directly using the time series as inputs/outputs for the machine learning model poses significant challenges (Liu et al., 2022). Therefore, an alternative approach is required to effectively handle this complexity. As indicated in Fig. 1, the signals generated from both the XBSB and XBNH models are first converted to the time-frequency domain using the Morlet wavelet transform method. This transformation generates scalograms that visually represent the frequency content of the wave runup simulations as a function of time. These scalograms serve as a valuable visual representation, enabling an easy identification of interesting features and patterns within the data efficiently. The obtained scalograms are further transformed into images and used by the machine learning model. Specifically, the scalograms (images) derived from the XBSB simulations are utilized as the inputs for the machine learning model, while the corresponding scalograms (images) generated from the XBNH simulations are treated as the outputs of the model. In this study, a Conditional Generative Adversarial Network (cGAN) is employed to establish a mapping between the image representations of the XBSB-based scalograms and the XBNH-based scalograms. The cGAN model is trained to generate XBNH-based scalograms, which are subsequently used to reconstruct the time-series wave runup results through the inverse wavelet transform as indicated in Fig. 1. A detailed discussion of the cGAN model is provided in the subsequent section.

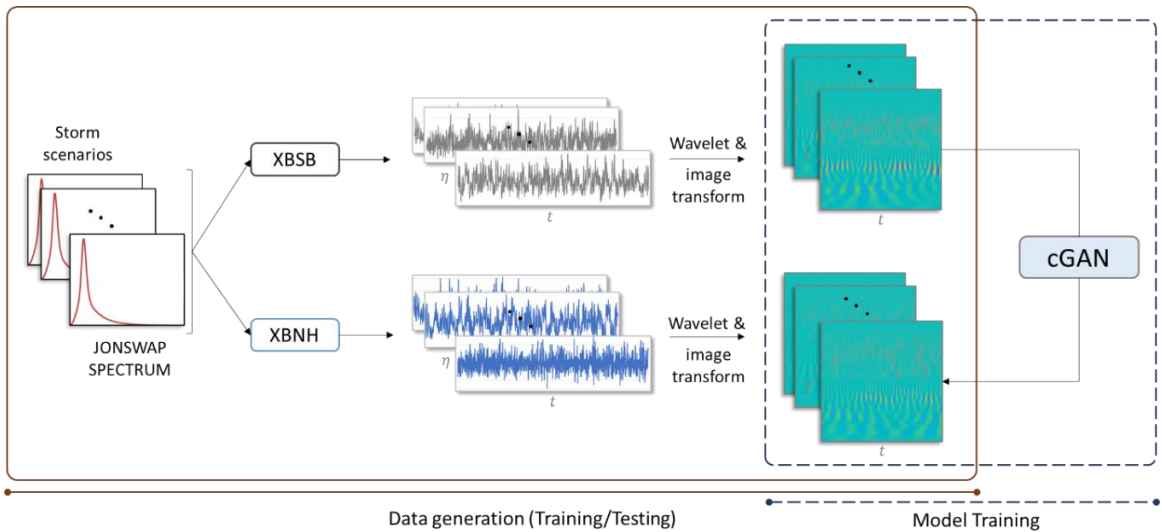

**Fig. 1.** Schematic framework of the methodology for data generation and model training

### 3.2. Machine learning methods

In this study, a Conditional Generative Adversarial Network (cGAN) is employed to establish a mapping between the XBSB-based image and its corresponding XBNH-based image, representing the scalograms (Isola et al., 2017). The cGAN model is a type of GAN which generates data (similar to GAN) that satisfies certain conditions (Mirza and Osindero, 2014). A typical GAN model consists of two neural networks, namely a generator and a discriminator. The generator is responsible for creating new (synthetic) data that should resemble the real data, while the discriminator is responsible for determining whether the data from the generator is real or fake.

The two networks are trained in an adversarial manner, engaging in a constant competition to outperform each other (Goodfellow et al., 2014). The key difference between the GAN and cGAN network lies in the input and output mechanisms. In a standard GAN model, the generator network takes random noise as input and generates synthetic data as output and the discriminator network tries to distinguish between real and fake data. On the other hand, a cGAN model introduces the concept of conditional input. Specifically, in addition to random noise, the generator in a cGAN also takes additional conditioning information as input which could be a specific label or some other form of auxiliary information that provides additional context for generating the output. Similarly, the discriminator, receives both real and fake data along with the same conditioning information. This conditioning mechanism enhances the control over the generated samples, making cGAN a promising approach for various applications, including image-to-image mapping. The typical architecture of the cGAN model is depicted in Fig. 2 (Isola et al., 2017; Ji et al., 2018).

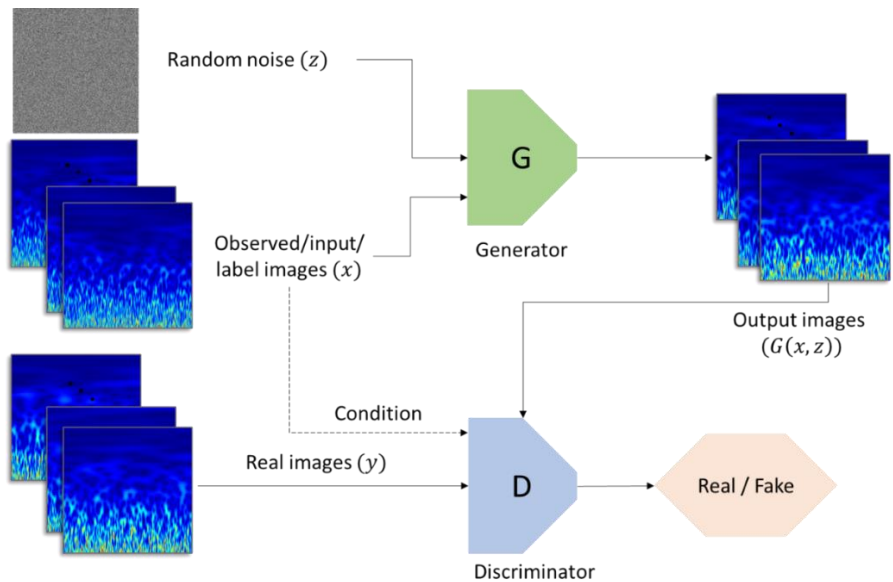

**Fig. 2.** Typical architecture of conditional GANs

The objective function of a cGAN model can be generally expressed in terms of the generator loss and the discriminator loss. While the generator loss aims to minimize the discrepancy between the generated samples and the real samples, the discriminator loss aims to maximize the discriminator's ability to correctly classify between real and generated samples (Chrysos et al., 2018). The objective loss function for the cGAN model can be expressed as:

$$\mathcal{L}_{cGAN}(G, D) = \mathbb{E}_{x,y}[log\, D(x,y)] + \mathbb{E}_{x,z}[\log(1 - D(x, G(x,z)))] \qquad (11)$$

where $x$ = input (observed) image; $y$ = output (real) image; $z$ = random noise vector; $G$ = generator; and $D$ = discriminator. The discriminator is trained to maximize this objective function, while the generator is trained to minimize it. In addition to the binary cross entropy loss term, an $L1$ loss term is added to the objective function to improve the results and decrease blurring (Larsen et al., 2016). This additional term affects essentially the generator to constrain it to be near the ground truth output and is expressed as:

$$\mathcal{L}_{L1}(G) = \mathbb{E}_{x,y,z}[\|y - G(x,z)\|_1] \tag{12}$$

Therefore, the total objective function can be expressed as:

$$G^* = arg \min_G \max_D \mathcal{L}_{cGAN}(G,D) + \lambda \mathcal{L}_{L1}(G) \tag{13}$$

where $\lambda$ = weighting parameter which penalizes the loss between the real image '$y$' and the generated image $G(x,z)$.

In this study, the generator learns a mapping from the XBSB-based scalograms '$x$' and random noise vector '$z$', to the corresponding XBNH-based scalograms '$y$'. On the other hand, the discriminator is trained to distinguish between the real pairs from fake pairs. From this perspective, the positive samples consist of pairs of XBSB-based scalograms '$x$' and their corresponding XBNH-based scalograms '$y$'. On the other hand, the negative samples are represented by pairs of XBSB-based scalograms '$x$' and XBNH-based scalograms generated by the generator network '$G(x,z)$'. While the first term on the right-hand side of Eq. (11) represents the loss when training the discriminator with positive samples, the second term represents the loss when training the discriminator with negative samples. The cGAN network contains a generator with a modified U-Net based architecture and a discriminator represented by a convolutional PatchGAN classifier which attempts to distinguish between the real pairs from fake pairs. The U-Net consists of an encoder and decoder with skip connections between them (Isola et al., 2017; Li and Wand, 2016; Long et al., 2015). The $L1$ loss is designed as a mean absolute error (MAE) between the generated image and the target image. More details about the model architecture will be provided in the case study section.

## 4. Illustrative Case Study

This section demonstrates the application of the proposed methodology to a simplified case study involving a 1D profile with varying depths. The generation of the required training and testing datasets will be first covered. Subsequently, the training process will be explained in detail. Finally, the trained model will be applied to various scenarios to assess its effectiveness.

### 4.1. Numerical model setup

The numerical simulations were configured using a 1D (cross-shore) profile with bathymetry resembling the laboratory experiment conducted in a large-scale wave flume by Demirbilek et al. (2007). The domain size extends approximately 30 m in the cross-shore direction. The depth of the 1D profile ranges from -0.5 m offshore to 0.4 m nearshore. The still water level (SWL) is set at 0.05 m as indicated in Fig. 3.

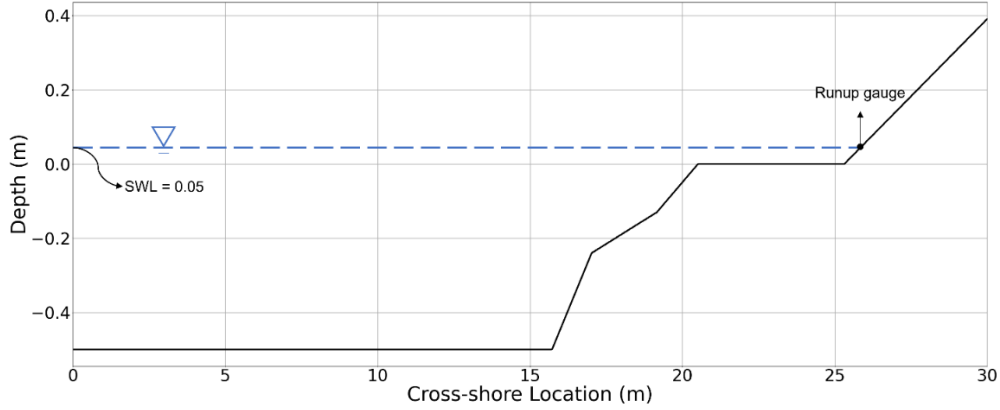

**Fig. 3.** Selected profile for simulation of time-series wave runup

A uniform grid size of 2.5 cm is utilized for XBNH, whereas for XBSB, the grid spacing varies from 5 cm offshore to 2.5 cm closer to the shoreline. The manning coefficient was set to 0.01 $\frac{s}{\sqrt[3]{m}}$ (Lashley et al., 2018). The numerical models are forced by a spectral condition represented by the JONSWAP spectrum. Six JONSWAP parameters are typically required by XBeach, namely the significant wave height ($H_{m0}$), peak frequency ($f_p$), peak enhancement factor ($\gamma$), main wave angle (*mainang*), directional spreading coefficient (*dsc*), and the highest frequency (*fnyq*). The generation of the storm scenarios was achieved by varying the first three JONSWAP parameters (i.e., $H_{m0}$, $f_p$, $\gamma$) while the remaining parameters *mainang*, *dsc* and *fnyq* were set to 270 *deg*, 1000 and 1 *Hz*, respectively for all of the simulations (Lashley et al., 2018). Specifically, a total of 100 experiments were conducted for each mode (i.e., XBNH and XBSB) to simulate 1800 seconds of wave runup with $0.05 \leq H_{m0}(m) \leq 0.085$, $0.55 \leq f_p(Hz) \leq 1$, and $1 \leq \gamma \leq 3.3$. It's worth noting that, in this case study, the model has been trained using a wide range of parameters for waves. Some of these parameters were originally used in laboratory and numerical studies conducted by the US Army Corps of Engineers to investigate runup conditions over Fringing Reefs (Demirbilek and Nwogu, 2007; Demirbilek et al., 2007). However, the model is designed to easily incorporate additional scenarios when needed, requiring less training effort, possibly through techniques such as transfer learning. Furthermore, the initial 150 seconds of each numerical simulation were considered as the spin-up time, while the subsequent 1650 seconds were used for the analysis. The spin-up time has been selected after analyzing the wave runup time-series for all simulated scenarios. This analysis indicated that a spin-up time of 150 seconds is appropriate across the different cases. On average, each simulation using the XBNH mode lasted approximately 5 minutes, whereas the simulations using the XBSB mode required less than 2 minutes and a half to complete. Using the generated database, the training of the proposed machine learning model can be conducted, as explained in the subsequent section.

## 4.2. Training process

As described in Section 3.1, the generated time series data from XBSB and XBNH undergo a series of transformations. Initially, they are converted into scalograms and subsequently transformed into images. The resulting spectrograms are structured as matrices with dimensions of 78 × 1650, representing the frequency and time content, respectively. These generated scalograms are then saved as RGB images with a pixel resolution of 1024 × 1024. To prepare the training set, a preprocessing step is applied. This involves introducing random jitter to the images. Specifically, the 1024 × 1024 images are resized to 1100 × 1100 and then randomly cropped back to their original size of 1024 × 1024. Additionally, a random mirroring technique is implemented by horizontally flipping the images. The images are then normalized in the [-1, 1] range for better data representation and effective model training.

    The generator employed in the cGAN model is based on a modified U-Net architecture, featuring a convolutional symmetric encoder-decoder structure. The encoder comprises 10 blocks, where each block consists of a Convolution/Batch-Norm/ReLU layer with $k$ filters denoted as C$k$. The selected encoder architecture can be represented as C64-C128-C128-C256-C256-C512-C512-C512-C512-C512. Similarly, the decoder consists of 9 blocks, each consisting of a Convolution/Batch-Norm/Dropout/ReLU layer with a dropout rate of 50% and $k$ filters denoted as CD$k$. The selected decoder architecture can be represented as CD512-CD512-CD512-CD512-CD256-CD256-CD128-CD128-CD64. All convolutions in the architecture utilize 4×4 spatial filters applied with a stride of 2. The ReLU activations in the encoder are leaky, with a slope of 0.2, while the ReLUs in the decoder are not leaky. Following the last layer in the decoder, a convolution is applied to map to the number of output channels (3 in this case representing the number of channels for RGB images) followed by a Tanh function to constrain the pixel values in the generated image in the range [-1,1]. To facilitate information flow and improve performance, skip connections are incorporated in the network architecture between encoding layers and the corresponding decoding layers, as depicted in Fig. 4. The output of the generator is an RGB image with a resolution of 1024×1024 pixels.

    The discriminator in the conditional Generative Adversarial Network (cGAN) model is designed to predict the likelihood of whether an image is real or a fake. It is constructed as a convolutional PatchGAN classifier, comprising four blocks of convolutional layers with ReLU activations and Batch normalization. The architecture of the discriminator can be represented as C64-C128-C256-C512. A distinctive feature of the PatchGAN classifier is its ability to determines whether each patch is real or fake and outputs the final result by averaging the responses of all the patches. It is important to note that Batch normalization is not applied to the first C64 layer in the discriminator architecture. All convolutions in the discriminator employ 4×4 spatial filters applied with a stride of 2. The ReLU activations in the discriminator network are leaky, with a slope of 0.2. After the last layer of the discriminator, a convolutional layer is employed to map the output to a 1-dimensional representation, followed by a Sigmoid function. The architecture of the selected cGAN model, including both the generator and discriminator is shown in Fig. 4.

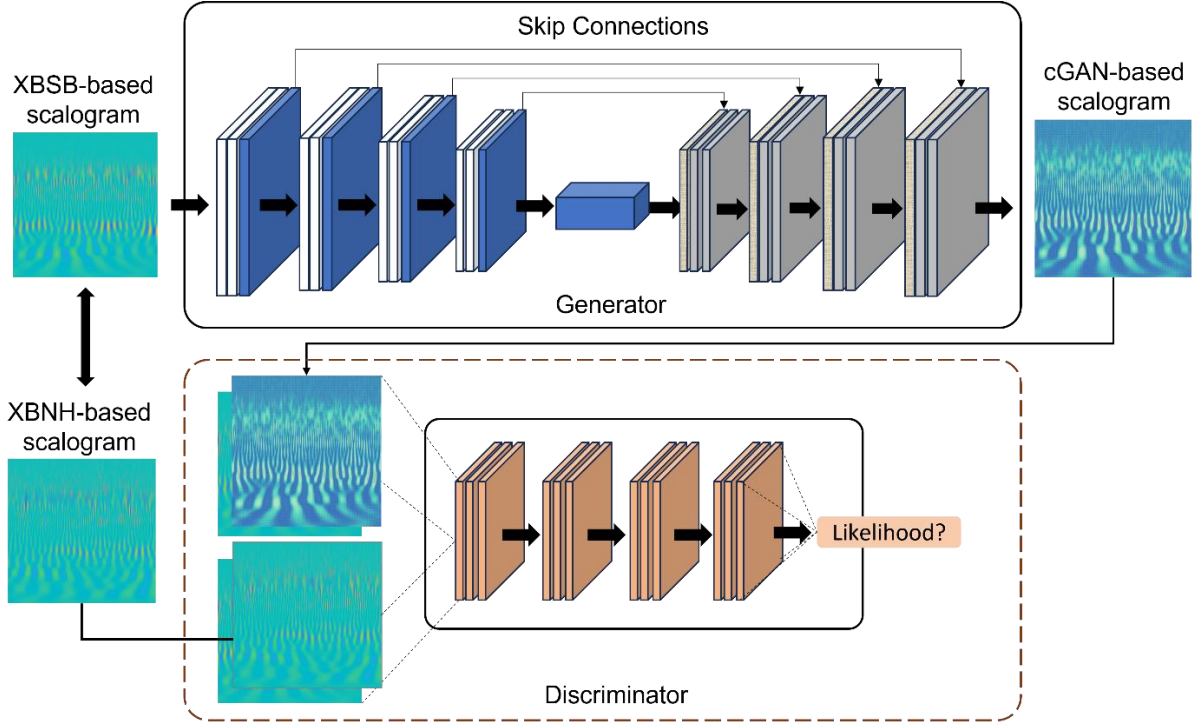

**Fig. 4.** Architecture of the proposed cGAN model

The remaining hyperparameters were determined through a trial-and-error approach. The training process utilized the batch size of one with the Adam optimization technique, employing a learning rate of 0.0002 and a momentum parameter $\beta_1 = 0.5$. The weights of both the generator and discriminator were initialized using a Gaussian distribution with a mean of 0 and a standard deviation of 0.02. The weighting parameter $\lambda$ of Eq. (13) was set to a large value of 100 to encourage the model to produce more hard negative samples. A total of 11,000 steps (epochs) were selected to train the cGAN model. The decision to use 11,000 training epochs was driven by the need for both the generator and discriminator to be effectively trained concurrently, as will be highlighted in the training process. The implementation of the entire framework is based on pix2pixGAN project (Isola et al., 2017). To evaluate the performance of the trained model, the dataset was randomly divided into a training set comprising 90% of the data and a separate test set comprising the remaining 10%.

In general, a well-trained conditional Generative Adversarial Network (cGAN) model exhibits strong performance from both the generator and discriminator. The generator should be capable of generating realistic images that are indistinguishable from real images, while the discriminator should accurately differentiate between real and generated images. However, training a cGAN model can be challenging, and simply increasing the number of training epochs does not necessarily guarantee better quality. Monitoring the loss values can provide insights into the training progress. It is important to note that GAN models typically do not converge but instead reach an equilibrium where neither network dominates the other. Therefore, monitoring this equilibrium is crucial to ensure the optimal performance of both networks. Although there is no definitive threshold for what constitutes a well-trained cGAN model, a general guideline is to assess the discriminator loss around 0.8 and the generator loss within the range of 0.5 to 2.0 (Brownlee, 2019). However, visual inspection of the generated samples, evaluation metrics, and domain-specific considerations are also important factors in determining the model's overall quality.

By examining the trained binary cross-entropy losses and visually inspecting the quality of the generated images, it can be concluded that the cGAN model has attained a satisfactory level of performance. Specifically, the generator loss falls within the range of 0.5 to 1.2, indicating effective generation capabilities. The discriminator loss fluctuates within an acceptable range, with values even reaching close to 0.85, as demonstrated in Fig. 5. It should be noted that the variability in the loss function is expected since both the generator and discriminator are in constant competition during the training process. This variability could potentially be reduced by identifying an optimal model architecture and parameters using advanced techniques like Bayesian optimization, instead of relying solely on trial-and-error methods, and by training the model with a larger dataset.

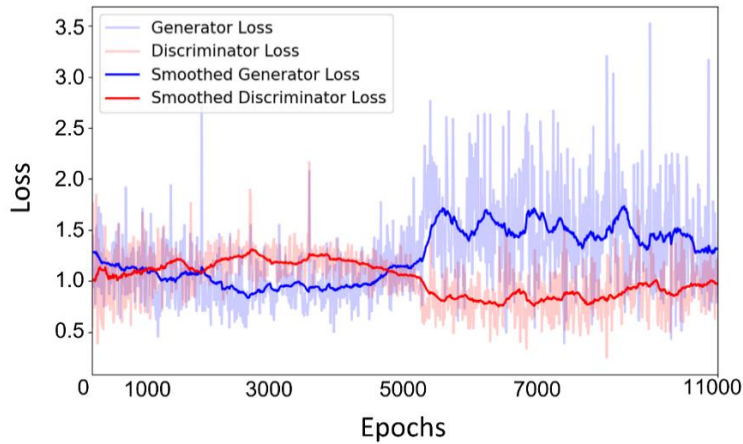

**Fig. 5.** Performance of the training process of the generator and the discriminator models [the smoothed curves correspond to a 32-step moving average]

Similarly, the L1 loss term in Eq. (13) is illustrated in Fig. 6, showing a consistent decreasing trend. This decreasing pattern indicates improved simulation results and a reduction in blurring, resulting in sharper images. In the next section, multiple examples of image inspection will be presented to provide a comprehensive assessment of the model's performance. These examples will serve as a means to evaluate the generated images and gain a deeper understanding of the model's capabilities.

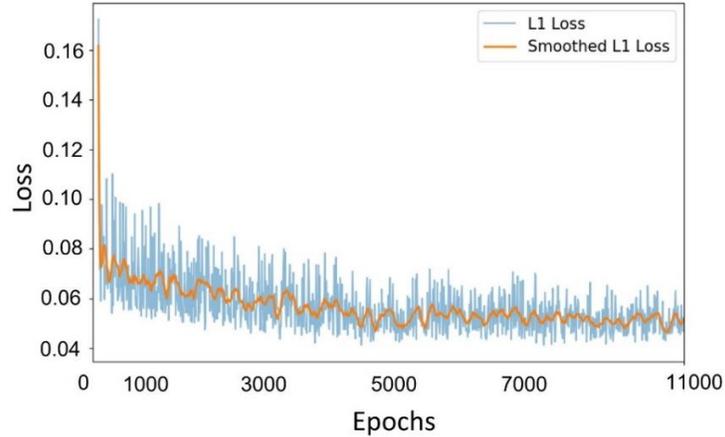

**Fig. 6.** L1 loss of the cGAN generator [the smoothed curve corresponds to a 32-step moving average]

### 4.3. Application

After training the cGAN model, it can be utilized to predict wave runup time histories. Specifically, the storm scenarios are selected from the testing set and employed as boundary conditions for the selected basin, as depicted in Fig. 1. The JONSWAP spectrum is utilized to define these storm scenarios as explained in Sect 4.1. Next, the XBSB model is executed with these storm scenarios to generate low-fidelity wave runup time histories. The simulation results obtained from the XBSB model are then transformed into scalograms using the Morlet wavelet transform. These scalograms are further converted into RGB images, which are subsequently fed into the trained cGAN model. The cGAN model predicts high-fidelity images that represent enhanced wave runup simulations. To retrieve the time history of the wave runup, the generated high-fidelity images are transformed back into scalograms. An inverse wavelet transform is then applied to obtain the corresponding time history. Both the generated images and the time history of the wave runup are compared with those generated by the XBNH model. This comparison serves as an assessment of the cGAN model's performance in generating high-fidelity wave runup simulations. Fig. 7 provides an illustration of the proposed methodology, outlining the steps involved in generating high-fidelity wave runup simulations using the cGAN model.

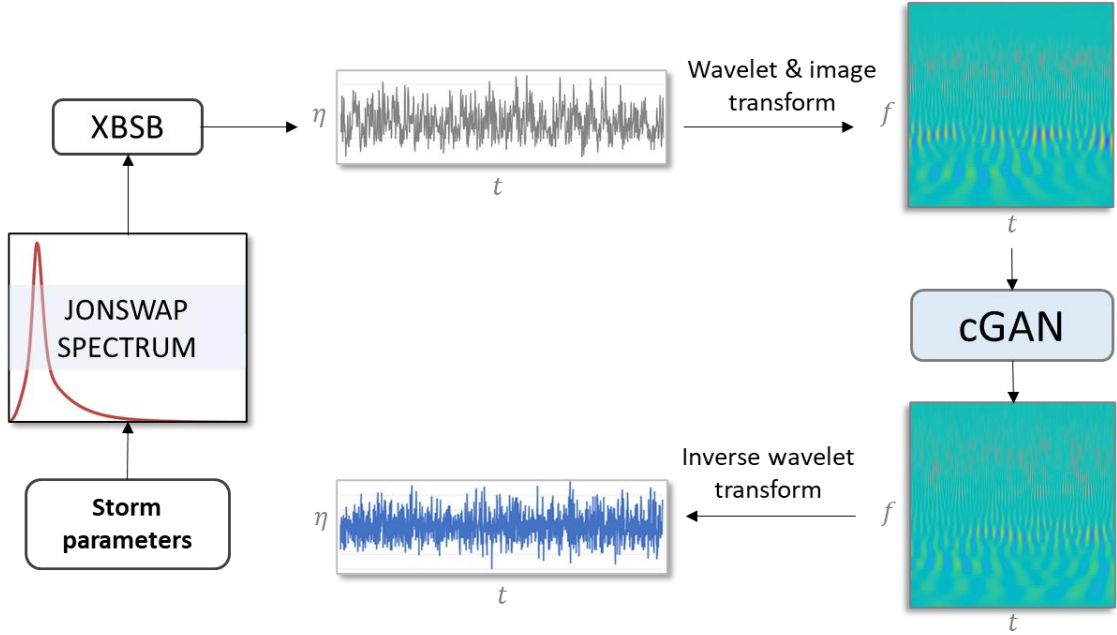

**Fig. 7.** Application of the proposed model for the prediction of high-fidelity time-series wave runup

Table 1 presents a summary of the three selected storm scenarios from the testing set, characterized by the JONSWAP spectrum.

**Table 1.** Storm parameters for time-series wave runup prediction

| Parameter | $H_{m0}$ (m) | $f_p$ (Hz) | $\gamma$ |
|---|---|---|---|
| First scenario | 0.056 | 0.918 | 1.418 |
| Second scenario | 0.065 | 0.800 | 2.022 |
| Third scenario | 0.084 | 0.554 | 3.276 |

The XBSB-based scalograms, generated using the storm parameters listed in Table 1, were inputted into the cGAN model. This resulted in the generation of image scalograms, as shown in the third column of Fig. 8. A thorough visual inspection reveals a good agreement between the generated scalograms and the XBNH-based scalograms (second column of Fig. 8). However, some discrepancies can still be identified in the figure, which are attributed to the complex nature of the system under study, characterized by its high-frequency and time-dependent content. The mean squared errors (MSE) between the cGAN-based and the XBNH-based scalograms are 1.06e-06, 1.26e-06, and 2.37e-06 for the first, second, and third scenarios, respectively. On the other hand, the MSE between the XBSB-based and the XBNH-based scalograms are 1.91e-05, 2.54e-05, and 5.21e-05 for the first, second, and third scenarios, respectively. The obtained results support the conclusion that the model was well trained.

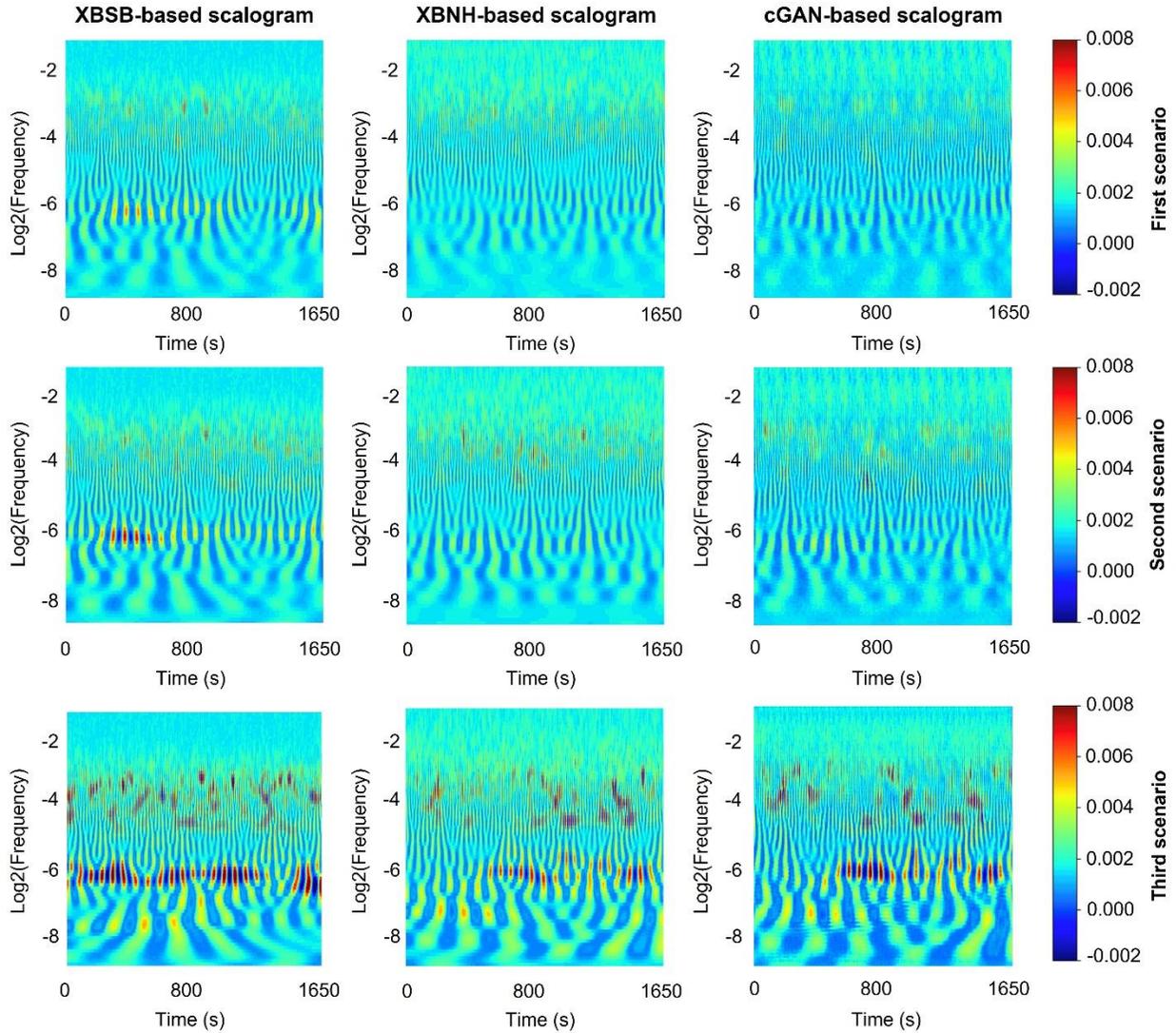

**Fig. 8.** Predicted scalograms (right column) given the XBSB-based scalograms (left column) as the inputs to the cGAN model of the first scenario (first row), second scenario (second row), and the third scenario (third row) along with the XBNH-based scalograms (ground truth, middle column)

Figure 9 shows the wave runup time histories corresponding to the three selected scenarios presented in Table 1. By comparing the output of the proposed cGAN model with the results based on XBNH, it can be concluded that a good agreement has been achieved for the simulation of the highly fluctuating wave runup process. Moreover, a zoomed plot is provided as an illustration for the third storm scenario, demonstrating a relatively good correlation between the time histories generated by c-GAN model and those derived from XBNH.

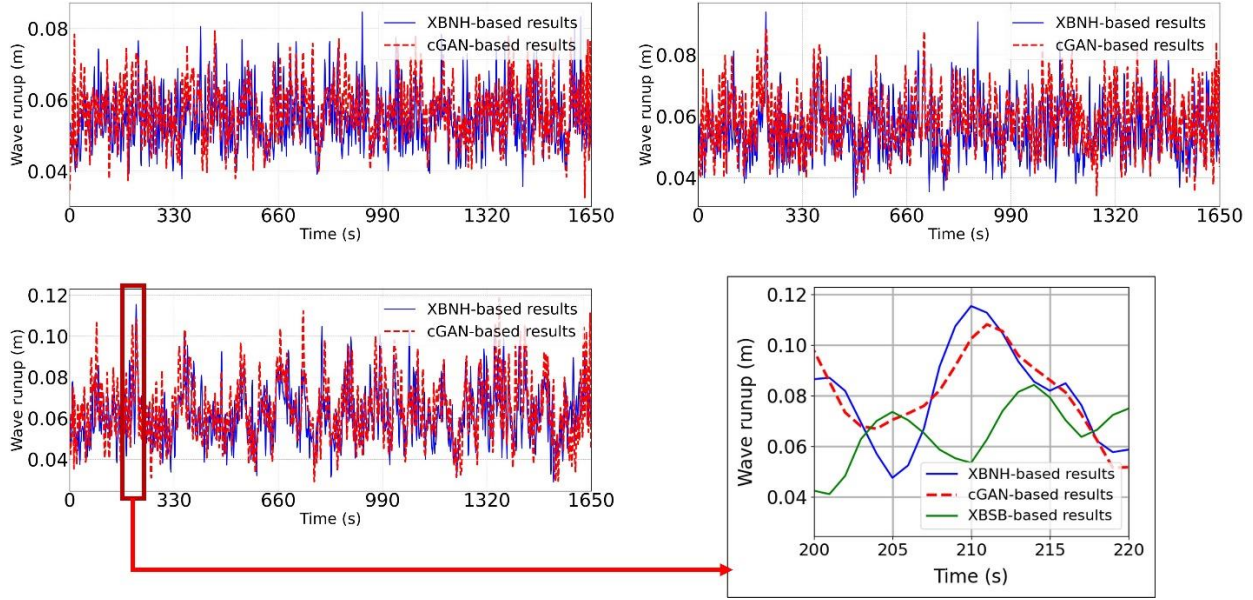

**Fig. 9.** Time-series wave runup based on XBNH and cGAN predictions for the first (top left), second (top right), and third (bottom) scenarios

In order to evaluate the accuracy of the reconstructed signals, several metrics were used. The mean, and standard deviation ($\sigma$) were calculated for the signals generated based on cGAN, XBNH, and XBSB. Furthermore, the MSE and the mean absolute error (MAE) were reported in Table 2.

**Table 2.** Comparison of the wave runup results generated by XBNH (ground truth), XBSB and c-GAN for the three selected storm scenarios

| Metrics | XBNH | | cGAN | | | | XBSB | | | |
|---|---|---|---|---|---|---|---|---|---|---|
| | Mean (m) | $\sigma$ (m) | Mean (m) | $\sigma$ (m) | MSE (m$^2$) | MAE (m) | Mean (m) | $\sigma$ (m) | MSE (m$^2$) | MAE (m) |
| 1st scenario | 0.055 | 0.007 | 0.056 | 0.007 | 9.16e-5 | 0.0001 | 0.053 | 0.008 | 1.18e-4 | 0.008 |
| 2nd scenario | 0.055 | 0.008 | 0.053 | 0.008 | 1.12e-4 | 0.008 | 0.056 | 0.010 | 1.57e-4 | 0.010 |
| 3rd scenario | 0.061 | 0.013 | 0.058 | 0.012 | 1.64e-4 | 0.010 | 0.059 | 0.014 | 3.23e-4 | 0.015 |

Based on the simulation results, it can be concluded that the mean, and standard deviation values obtained from the cGAN model are nearly identical to those from the XBNH-based simulations. Furthermore, the low values of MSE and MAE metrics suggest that the cGAN model exhibits good accuracy in reproducing the wave runup time history. Table 2 also includes the low-fidelity XBSB results, which evidently exhibit higher errors when compared to the proposed model.

## 5. Discussion

Wave runup is a complex process characterized by its inherent nonlinearity and transient, nonstationary nature. Therefore, predicting the temporal evolution of wave elevation remains highly challenging, even with the most advanced machine learning techniques. As a result, many current applications of ML for predicting wave runup focus on predicting specific statistical measures, such as $R_{max}$ and $R_{2\%}$. This study introduces a new model based on the cGAN technique, which aims to predict the time-history of wave runup using available data from the physics-based XBSB model as input. Specifically, the proposed model maps the image representation of wave runup from XBSB to the corresponding image from XBNH. Utilizing the scalograms and their corresponding images allows the model to effectively identify and separate the most significant frequency components, such as low and high frequencies, thereby facilitating the training process.

An essential part of the model training involved transforming the time-series wave runup into scalograms and then into RGB images. This process was carefully examined to assess any errors introduced by this direct transformation and its inverse. In general, the induced errors were negligible. For instance, in the case of a benchmark data example (third scenario), the error has been investigated at two levels: firstly, in relation to the initial transformation from time history signals to scalograms, and secondly, in relation to the subsequent transformation from the scalograms to the RGB images. In the first transformation, the error was calculated by comparing a given time history signal with the one obtained based on the scalogram using the inverse wavelet transform. The resulting MSE value between the two signals is 0.000434 (for the normalized signals). In the second transformation, the error was calculated by comparing a given scalogram with the one obtained based on the corresponding RGB image. The resulting MSE value between the two scalograms is 1.37e-08. This indicates that the proposed approach does not alter the original signals, as they can be retrieved almost identically using the inverse transforms. On the other hand, despite training the model on a relatively large parameter range for waves (Demirbilek and Nwogu, 2007; Demirbilek et al., 2007), there are still other scenarios that could be important to explore. Therefore, these scenarios could potentially be included in the training set. Similarly, although the model can predict wave runup under various storm conditions, its applicability is restricted to a single coastal profile (Fig. 3). With a new basin configuration, it's necessary to retrain the model to accommodate these changes. Several techniques like transfer learning can expedite this process by leveraging knowledge acquired from prior training on similar tasks or datasets. This approach can significantly reduce the training time and resources required for the updated model.

Training a cGAN model is also very challenging since both the generator and discriminator are in constant competition during the training process. As mentioned before, cGAN models typically do not converge but instead reach an equilibrium where neither network dominates the other. Hence, it's important to monitor the training process to assess the discriminator and generator's performance in line with some established guidelines (Sect. 4.2). Additionally, a visual inspection of the generated samples during the training process should complement these evaluations. Visual inspection provides a qualitative assessment that can offer valuable insights into the model's performance and the quality of the generated samples. During the training process, numerous hyperparameters were fine-tuned to find the optimal model configuration through a trial-and-error method. However, more advanced techniques like Bayesian optimization could be employed to identify the best parameters. This approach could potentially enhance the training of the cGAN model by efficiently exploring the hyperparameter space and finding configurations that lead to improved performance.

Following the model training, two critical metrics were examined: accuracy and efficiency. These metrics are essential for evaluating the model's performance and its computational effectiveness in producing accurate results. The analysis of the training and testing results (Sect. 4.2 and 4.3) revealed that the cGAN model can generally make accurate predictions of time-dependent wave runup. However, some discrepancies were noted, as discussed earlier, which could potentially be addressed to further improve the model's predictive capabilities. After the cGAN model has been trained, it can instantaneously generate the wave runup predictions. Yet, the cGAN model necessitates the use of low-fidelity simulations from XBSB mode as input to produce the corresponding high-fidelity simulations, leading to a total simulation time comparable to XBSB. Generally, for the selected profile (Fig. 3), the entire process takes approximately 2.5 minutes for a single prediction (which corresponds to a wave runup simulation over 30 minutes). In comparison, the same simulation would take nearly 5 minutes using the high-fidelity XBNH mode. Therefore, the proposed model reduces the time required for high-fidelity simulation by almost half. In real-time applications, this effect will be even more significant because it is crucial to accommodate uncertain input parameters, which would necessitate more simulations for the same storm condition. Similarly, in probabilistic and risk simulations, typically involving hundreds of thousands of scenarios (e.g., Snaiki and Wu, 2020b, 2022), the proposed approach will result in significant time savings. For example, with 1,000 simulations, the proposed model would take approximately 1,000 * 2.5 minutes, totaling around 41.6 hours, compared to 1,000 * 5 minutes, or roughly 83.3 hours using the high-fidelity XBNH mode. This difference translates to saving almost 41 hours of simulation time. To further reduce the computational cost, it is important to select a different model instead of XBSB, which dictates the prediction time needed for the proposed model. This can be achieved by using empirical methods or even machine learning techniques capable of generating low-fidelity simulations.

## 6. Conclusion

In this study, a physics-informed machine learning model is proposed to efficiently and accurately simulate time-series wave runup. Specifically, a cGAN model is used to generate high fidelity-based XBNH wave runup from low fidelity-based XBSB wave runup. By leveraging the physics-based knowledge from the XBSB mode as input, this conditional mechanism enhances the control over samples generated by the cGAN model, consequently improving its performance in image-to-image mapping applications. The storm scenarios were implemented as the boundary conditions of the selected basin using a simplified representation based on the JONSWAP spectrum. The input/output images for the cGAN model were generated using the Morlet wavelet transform technique, resulting in scalograms that were subsequently saved as RGB images. Once the model was trained, the high-fidelity XBNH-based scalograms were predicted, which were then utilized for reconstructing the time-series wave runup based on the inverse wavelet transform. The mean squared errors (MSE) between the cGAN-based and the XBNH-based wave runup were minimal. For instance, for three scenarios that were selected from the testing set, the MSE values were 9.16e-5, 1.12e-4, 1.64e-4 for the first, second and third scenario, respectively which are lower than those generated by XBSB. Additionally, the mean and standard deviation values of the time-series wave runup signals obtained by the cGAN model closely match those from the XBNH-based results for the three selected scenarios. Furthermore, after training the cGAN model, it can instantly predict the time-series wave runup for a given storms scenario. As the proposed model relies on low-fidelity simulations from XBSB mode as input to generate the corresponding high-fidelity simulations, it leads to a total simulation time similar to XBSB, effectively reducing the time needed for high-fidelity simulation by nearly half. The obtained results indicate that the proposed model has the potential to serve as a tool for rapid risk assessment applications. Despite demonstrating a good accuracy and efficiency in simulating the time-series wave runup, some limitations have been identified. These are attributed to the complex nature of the system under study, characterized by its high-frequency and time-dependent content. Therefore, further refinement of the model is needed. Specifically, the training dataset should include various coastal bathymetries along with new storm conditions. Additionally, exploring more advanced techniques like Bayesian optimization to train the cGAN model could potentially enhance its performance by efficiently exploring the hyperparameter space and identifying configurations that result in improved results. Furthermore, in order to further decrease the computational cost, it is suggested to choose an alternative model other than XBSB, which determines the prediction time required for the proposed model, and this can be accomplished by employing empirical methods or even machine learning techniques capable of producing low-fidelity simulations.

**Declaration of Competing Interest**

The authors declare that they have no known competing financial interests or personal relationships that could have appeared to influence the work reported in this paper.

**Acknowledgements**

This work was supported by the Natural Sciences and Engineering Research Council of Canada (NSERC) [grant number CRSNG RGPIN 2022-03492].

De Beer, A. F., McCall, R. T., Long, J. W., Tissier, M. F. S., & Reniers, A. J. H. M. (2021). Simulating wave runup on an intermediate–reflective beach using a wave-resolving and a wave-averaged version of XBeach. *Coastal Engineering*, *163*, 103788.

Demirbilek, Z., Nwogu, O. G., & Ward, D. L. (2007). Laboratory study of wind effect on runup over fringing reefs, Report 1: data report.

Demirbilek, Z., & Nwogu, O. G. (2007). Boussinesq modeling of wave propagation and runup over fringing coral reefs, model evaluation report.

Di Luccio, D., Benassai, G., Budillon, G., Mucerino, L., Montella, R., & Pugliese Carratelli, E. (2018). Wave run-up prediction and observation in a micro-tidal beach. *Natural Hazards and Earth System Sciences*, *18*(11), 2841-2857.

Didier, D., Bernatchez, P., Boucher-Brossard, G., Lambert, A., Fraser, C., Barnett, R. L., & Van-Wierts, S. (2015). Coastal flood assessment based on field debris measurements and wave runup empirical model. *Journal of Marine Science and Engineering*, *3*(3), 560-590.

Didier, D., Bernatchez, P., Marie, G., & Boucher-Brossard, G. (2016). Wave runup estimations on platform-beaches for coastal flood hazard assessment. *Natural Hazards*, *83*, 1443-1467.

Fiedler, J. W., Smit, P. B., Brodie, K. L., McNinch, J., & Guza, R. T. (2018). Numerical modeling of wave runup on steep and mildly sloping natural beaches. *Coastal Engineering*, *131*, 106-113.

Goodfellow, I., Pouget-Abadie, J., Mirza, M., Xu, B., Warde-Farley, D., Ozair, S., ... & Bengio, Y. (2014). Generative adversarial nets. *Advances in neural information processing systems*, *27*.

Guimaraes, P. V., Farina, L., Toldo Jr, E., Diaz-Hernandez, G., & Akhmatskaya, E. (2015). Numerical simulation of extreme wave runup during storm events in Tramandaí Beach, Rio Grande do Sul, Brazil. *Coastal Engineering*, *95*, 171-180.

Harris, D. L., Rovere, A., Casella, E., Power, H., Canavesio, R., Collin, A., ... & Parravicini, V. (2018). Coral reef structural complexity provides important coastal protection from waves under rising sea levels. *Science advances*, *4*(2), eaao4350.

Hatzikyriakou, A., & Lin, N. (2018). Assessing the vulnerability of structures and residential communities to storm surge: An analysis of flood impact during Hurricane Sandy. *Frontiers in Built Environment*, *4*, 4.

Hedges, T. S., & Mase, H. (2004). Modified Hunt's equation incorporating wave setup. *Journal of waterway, port, coastal, and ocean engineering*, *130*(3), 109-113.

Hermawan, S., Bangguna, D., Mihardja, E., Fernaldi, J., & Prajogo, J. E. (2023). The Hydrodynamic Model Application for Future Coastal Zone Development in Remote Area. *Civil Engineering Journal*, *9*(8), 1828-1850.

Holman, R. A. (1986). Extreme value statistics for wave run-up on a natural beach. *Coastal Engineering*, *9*(6), 527-544.
23